\documentclass[a4paper,12pt]{article}

\usepackage{amssymb}
\usepackage{amsmath}
\usepackage[dvips]{graphicx}
\makeatletter
  
  \@addtoreset{equation}{section}
\makeatother
\makeatletter
\def\footnoteref#1{\def\@thefnmark{\ref{#1}}%
 \@footnotemark}
\makeatletter

\begin{document}

\begin{flushright}
OU-HET 377\\
hep-th/0101222\\
January 2001
\end{flushright}
\vspace*{1.5cm}
\begin{center}
{\Large {\bf Spontaneous Lorentz Symmetry Breaking 
\\ by Anti-Symmetric Tensor Field}} \\
\bigskip
Kiyoshi Higashijima\footnote{higashij@het.phys.sci.osaka-u.ac.jp} 
and Naoto Yokoi\footnote{yokoi@het.phys.sci.osaka-u.ac.jp}\\
\bigskip
{\small
Department of Physics,\\
Graduate School of Science, Osaka University,\\
Toyonaka, Osaka 560-0043, JAPAN
}
\end{center}
\bigskip
\bigskip
\bigskip

\begin{abstract}
We study the spontaneous Lorentz symmetry breaking in 
a field theoretical model in (2+1)-dimension, inspired by string theory.
This model is a gauge theory of an anti-symmetric
tensor field and a vector field (photon). 
The Nambu-Goldstone (NG) boson for 
the spontaneous Lorentz symmetry breaking is identified  
with the unphysical massless photon in the covariant 
quantization.
We also discuss an analogue of the equivalence theorem
between the amplitudes for emission or absorption of 
the physical massive anti-symmetric tensor 
field and those of the unphysical massless photon.
The low-energy effective action of the NG-boson is also discussed. 
\end{abstract}

\newpage

\section{Introduction}
Quantum field theories based on the Poincar\'e invariance, in
particular, the Lorentz invariance successfully describe elementary
particles below the weak scale energy ($\sim$ 100 GeV).
In a last few years, a possible type of the 
Lorentz \textit{non}-invariant extensions of 
the quantum field theories has been extensively studied.
These are the field theories on the space-time whose coordinates are
non-commutative, called the non-commutative field
theories\cite{Connes, Douglas-Hull, IKKT, Seiberg-Witten}.
The action of the non-commutative field theories can be 
constructed by replacing the product of fields in the action of
the ordinary field theory with the $\star$-product defined as
\begin{eqnarray}
f(x) \star g(x) \equiv e^{i\theta^{ij}\frac{\partial}{\partial
\xi^{i}}\frac{\partial}{\partial \eta^{j}}}
f(x+\xi)g(x+\eta)\bigl|_{\xi=\eta=0},
\end{eqnarray}
where $\theta^{ij}$ is a constant non-commutative parameter: 
$\left[x^{i}, x^{j}\right] = i \theta^{ij}$.
Thus the action explicitly contains the \textit{constant} 
anti-symmetric tensor $\theta^{ij}$, and 
the Lorentz invariance in (p+1)-dimension for $\textrm{p}\geq 2$
cannot be maintained.

String theory naturally provides the non-commutative field theories
as the world volume effective theories on D-branes\cite{Seiberg-Witten}:
the world volume effective theory of Dp-brane with a constant
background NS-NS B-field is equivalent to a (p+1)-dimensional
non-commutative field theory whose constant non-commutative parameter
$\theta^{ij}$ is given by the background NS-NS B-field $B_{ij}$.
In string theory the NS-NS B-field is indeed 
a dynamical field in closed string sector and thus the constant 
background field can be interpreted as the constant vacuum expectation 
value of the dynamical NS-NS B-field.
From this perspective, the Lorentz symmetry is \textit{spontaneously} 
broken by the constant vacuum expectation value of the second rank 
anti-symmetric tensor field. 

In this paper, based on this viewpoint, we discuss 
the spontaneous Lorentz symmetry breaking within the effective field
theory of the string theory. Concretely, we investigate the
Nambu-Goldstone boson for the Lorentz symmetry breaking in a field
theoretical toy model in (2+1)-dimension 
of a second rank anti-symmetric tensor field and
a vector field, which is inspired by the effective theory of
the string theory. 
We find that the NG-boson is an unphysical field and their amplitudes, 
however, provide the useful information about the physical amplitudes
of the model through the ``equivalence theorem''.
We also discuss the low-energy dynamics of the NG-boson from the
perspective of the nonlinear realization of the Lorentz symmetry. 

This paper is organized as follows.
In the next section, we introduce the gauge invariant model 
of a second rank anti-symmetric tensor field and 
a vector field and discuss the covariant canonical quantization of 
the model. 
In section 3, the vacuum of the model where the anti-symmetric 
tensor field has a constant vacuum expectation value is discussed 
and also the Nambu-Goldstone boson for the spontaneous Lorentz
symmetry breaking is studied in detail.
In section 4, a possible perturbation of the model is discussed
and the equivalence theorem between the physical amplitudes and the
amplitudes of the unphysical NG-boson is also argued.
In section 5, some related problems are discussed 
and the relation to the non-commutative field theories is speculated.
\section{A toy model for field theory of ${\boldsymbol B_{\mu \nu}}$ and 
${\boldsymbol A_{\mu}}$}
In this section we discuss the covariant canonical quantization of a 
toy model for the gauge invariant 
field theory of an second rank anti-symmetric 
tensor field $B_{\mu \nu}$ coupled with a vector field $A_{\mu}$
(photon) in (2+1)-dimension.   

\subsection{Canonical quantization of the model}

The action of the toy model is given by\footnote{The metric is
$\eta_{\mu \nu}=\eta^{\mu \nu}=\textrm{diag.}(+1,-1,-1)$.}
\begin{eqnarray}
S = \int d^{3}x \Bigl(\frac{1}{12 m^2} \left(H_{\mu \nu \rho}\right)^2
-\frac{1}{4}\left(F_{\mu \nu}-B_{\mu \nu}\right)^{2}\Bigr),
\label{gauge inv. action}
\end{eqnarray}
where
\begin{eqnarray}
H_{\mu \nu \rho}=\partial_{\mu} B_{\nu \rho} + \partial_{\rho} 
B_{\mu \nu} + \partial_{\nu} B_{\rho \mu}, \quad 
F_{\mu \nu}=\partial_{\mu} A_{\nu} - \partial_{\nu} A_{\mu},
\end{eqnarray}
and $m$ is a parameter with dimension of mass.
This action is inspired by string theory
\footnote{In fact, this type of action appears in various contexts of string
theory\cite{Kalb-Ramond, Cremmer-Scherk, Callan, Green}.}.
Indeed, the first and second term in (\ref{gauge inv. action}) are 
the same form as the leading term of the
effective action of $B_{\mu \nu}$,  
which is a massless mode of closed string, 
and the leading term of the Dirac-Born-Infeld (DBI) action
of D-brane world volume effective theory, which is the effective action of
the open string sector, in $\alpha^{'}$-expansion\cite{Polchinski}.
%We note that $m^{2}$ occupies a same place as $\alpha^{'}$ in this
%correspondence.
%\footnote{More concretely, the world volume theory
%of a flat D2-brane in string theory compactified on $T^{23}$,
%which is the transverse space of the brane are expected to be given by
%(\ref{gauge inv. action})}.

The action (\ref{gauge inv. action}) is invariant under the gauge
transformation:
\begin{eqnarray}
\delta B_{\mu \nu}(x) &=& \partial_{\mu} \Lambda_{\nu}(x)-
\partial_{\nu} \Lambda_{\mu}(x), \nonumber\\
\delta A_{\mu}(x) &=& \Lambda_{\mu}(x) + \partial_{\mu} \Lambda(x),
\label{gauge sym.}
\end{eqnarray}
where $\Lambda_{\mu}(x)$ and $\Lambda(x)$ are 1-form and scalar gauge
functions respectively.
Because of this gauge invariance, the system described by the action
(\ref{gauge inv. action}) is a singular (constrained) system. 
Thus, for the canonical quantization, one must introduce gauge fixing terms.
Since we want to discuss the spontaneous Lorentz symmetry breaking
in the sequel, we must take a Lorentz invariant gauge fixing terms.  
We introduce the following gauge fixing terms:
\begin{eqnarray}
S_{\textrm{gf}} = \int d^{3}x \Bigl(C^{\nu}\partial^{\mu}B_{\mu \nu} - 
B\partial_{\mu} A^{\mu} - C \partial_{\mu} C^{\mu} \Bigr), 
\label{landau gauge}
\end{eqnarray}
where $B(x)$ is the Nakanishi-Lautrup (NL) B-field for the vector field 
and $C_{\mu}(x)$ and $C(x)$ are the counterparts 
for the anti-symmetric tensor gauge field\cite{Nakanishi1, Nakanishi2}. 
These gauge fixing terms are the analogues of the Landau gauge in quantum
electrodynamics (QED). 

Although the canonical quantization of the model in the BRST formalism 
can be carried out, we make the canonical quantization in 
the NL formalism\cite{Nakanishi1, Nakanishi2} 
for simplicity\footnote{In the BRST formalism, 
ghost and anti-ghost fields are introduced in addition. However,
since this action is a quadratic action with abelian gauge symmetry,
ghost and anti-ghost fields are free fields and decouple}.

The gauge fixed action is given by (\ref{gauge inv. action}) and 
(\ref{landau gauge}):
\begin{eqnarray}
S_{\textrm{total}} = \int d^{3}x~{\cal L}_{\textrm{total}}(x) 
\hspace{-0.2cm}&=&\hspace{-0.2cm}\int d^{3}x \biggl(\frac{1}{12 m^2} 
\left(H_{\mu \nu \rho}\right)^2 -\frac{1}{4}
\left(F_{\mu \nu}-B_{\mu \nu}\right)^{2}\biggr.\nonumber\\
&& \biggl.\hspace{0.6cm}+~~C^{\nu}\partial^{\mu}B_{\mu \nu} - 
B\partial_{\mu} A^{\mu} - C \partial_{\mu} C^{\mu}\biggr).
\label{total action}
\end{eqnarray}    
The equations of motion derived from (\ref{total action}) for each
field become as follows.
\begin{eqnarray}
B_{\mu \nu} &:& \frac{1}{m^2} \partial_{\rho}H^{\rho \mu \nu} + \left(B^{\mu
\nu}-F^{\mu \nu}+C^{\mu \nu}\right) = 0, \label{eq. of B}\\
A_{\mu} &:& -\partial_{\rho}\left(F^{\rho \mu}-B^{\rho \mu}\right) -
\partial^{\mu} B = 0,\label{eq. of A}\\
C_{\mu} &:& -\partial_{\rho}B^{\rho \mu}-\partial^{\mu} C = 0,
\label{eq. of C}\\
B &:& \partial_{\mu} A^{\mu} = 0, \label{eq. of NLB}\\
C &:& \partial_{\mu} C^{\mu} = 0,\label{eq. of NLC}
\end{eqnarray}
where $C_{\mu \nu} = \partial_{\mu} C_{\nu} - \partial_{\nu} C_{\mu}$.   
Actually, by combining these equations, 
one can find free field equations of each field:
\begin{eqnarray}
&&\square^2 A_{\mu} = 0, \quad \square~(\square + m^2) B_{\mu \nu} =
0,\nonumber \\
&&\square^2 C_{\mu} = 0, \quad \square B = 0, \quad \square C = 0.
\label{free eq. of motion} 
\end{eqnarray}
Thus this model is essentially a free field theory and can be
quantized completely. 
Note that the anti-symmetric tensor field $B_{\mu \nu}$ is a mixture
of massive and massless components.
Following the procedure in \cite{Nakanishi1, Nakanishi2},
the three-dimensional commutation relations can be calculated
by using the equal-time commutation relations,
\begin{eqnarray}
&&\left[\phi_{I}({\boldsymbol x},t),\phi_{J}({\boldsymbol y},t)\right]=0,\qquad
\left[\pi^{I}({\boldsymbol x},t),\pi^{J}({\boldsymbol y},t)\right]=0,
\nonumber\\
&&\left[\phi_{I}({\boldsymbol x},t),\pi^{J}({\boldsymbol y},t)\right] = i
\delta^{J}_{I} \delta^{2}({\boldsymbol x}-{\boldsymbol y}), \quad \textrm{where}~~
\pi^{I}({\boldsymbol x},t)\equiv\frac{\partial {\cal L}_{\textrm{total}}(x)}
{\partial \dot{\phi}_{I}({\boldsymbol x},t)}.
\end{eqnarray}
Here we abbreviate various field as $\phi_{I}({\boldsymbol x},t)$, where 
$I$ denotes various indices. The explicit forms of the non-vanishing 
three-dimensional commutation relations are\footnote{The 
equal-time commutation relations are obtained by setting
$x^{0}=y^{0}$ in the three-dimensional commutation relations.} 
\begin{eqnarray}
\left[B(x), A_{\mu}(y)\right]\hspace{-0.2cm}&=&\hspace{-0.2cm}
\left[C(x), C_{\mu}(y)\right] = -i
\partial_{\mu}^{x}D(x-y), \nonumber\\ 
\left[A_{\mu}(x),A_{\nu}(y)\right]\hspace{-0.2cm}&=&\hspace{-0.2cm}
\left[C_{\mu}(x),A_{\nu}(y)\right]=-i \eta_{\mu \nu} D(x-y) + 
i \partial^{x}_{\mu} \partial^{x}_{\nu} E(x-y),\nonumber\\
\left[C_{\mu}(x),B_{\nu \rho}(y)\right]\hspace{-0.2cm}&=&
\hspace{-0.2cm}-i \left(\eta_{\mu \nu}
\partial_{\rho}^{x}-\eta_{\mu \rho}\partial_{\nu}^{x}\right)D(x-y),
\nonumber\\
\left[B_{\mu \nu}(x),B_{\rho \sigma}(y)\right]\hspace{-0.2cm}&=&
\hspace{-0.2cm}i \Bigl(\eta_{\mu \rho}\partial_{\nu}^{x}
\partial_{\sigma}^{x}-\eta_{\mu \sigma}\partial_{\nu}^{x}\partial_{\rho}^{x}
-\eta_{\nu \rho}\partial_{\mu}^{x}\partial_{\sigma}^{x}
+\eta_{\nu \sigma}\partial_{\mu}^{x}\partial_{\rho}^{x} \Bigr.
\nonumber\\
\hspace{-0.2cm}&&\hspace{-0.2cm}\Bigl. +~m^2
\left(\eta_{\mu \rho}\eta_{\nu \sigma}-\eta_{\mu \sigma}\eta_{\nu
\rho}\right)\Bigr)\Delta(x-y:m^2) \nonumber\\
\hspace{-0.2cm}&&\hspace{-0.2cm}-i\Bigl(\eta_{\mu \rho}\partial_{\nu}^{x}
\partial_{\sigma}^{x}
-\eta_{\mu \sigma}\partial_{\nu}^{x}\partial_{\rho}^{x}
-\eta_{\nu \rho}\partial_{\mu}^{x}\partial_{\sigma}^{x}
+\eta_{\nu \sigma}\partial_{\mu}^{x}\partial_{\rho}^{x}\Bigr)D(x-y),
\label{three-dimensional commutators}
\end{eqnarray}
where
\begin{eqnarray}
\Delta(x:m^2)&\equiv&\frac{1}{(2\pi)^2 i}\int d^{3}k~
\epsilon(k_{0})\delta(k^2-m^2)e^{-ikx},
\quad D(x)\equiv\Delta(x:m^2=0),\\
E(x)&\equiv&\frac{1}{(2\pi)^2 i}\int d^{3}k~\epsilon(k_{0})
\delta^{'}(k^2)e^{-ikx},\qquad
\square E(x) = D(x).
\end{eqnarray}

In order to quantize the model consistently, we require the physical
state conditions analogous to the ordinary QED in the NL
formalism\cite{Nakanishi1, Nakanishi2}.
We define the \textit{physical state} through the physical state conditions:
\begin{eqnarray}
C_{\mu}^{(+)}(x)|\textrm{phys}\rangle = 0, \quad 
B^{(+)}(x)|\textrm{phys}\rangle = 0, \quad 
C^{(+)}(x)|\textrm{phys}\rangle = 0,
\label{phys. cond. of field} 
\end{eqnarray}
where $\phi_{I}^{(+)}(x)$ means the positive energy part of
$\phi_{I}(x)$. In the gauge (\ref{landau gauge}), as seen from
(\ref{free eq. of motion}), $C_{\mu}(x)$ is a dipole field.
Although the separation between the positive and negative energy part
of a dipole field is a non-trivial problem, the cut-off procedure 
is known to give the well-defined separation as is found  
in the next subsection\cite{Nakanishi2, Nakanishi3}.
Thus the physical state conditions (\ref{phys. cond. of field})
are well-defined. 
\subsection{The physical spectrum} 
In order to find the spectrum of the model, we define the creation and 
annihilation operators of each field.
The annihilation operators are defined by the Fourier transforms:
\begin{eqnarray}
C^{(+)}(x)=\frac{1}{2 \pi}\int d^{3}k~\theta(k_{0})e^{-ikx}b(k),
\nonumber\\
B^{(+)}(x)=\frac{1}{2 \pi}\int d^{3}k~\theta(k_{0})e^{-ikx}c(k),
\nonumber\\
C_{\mu}^{(+)}(x:\epsilon)=\frac{1}{2 \pi}\int d^{3}k~
\theta(k_{0}-\epsilon)e^{-ikx}c_{\mu}(k), \nonumber\\
A_{\mu}^{(+)}(x:\epsilon)=\frac{1}{2 \pi}\int d^{3}k~
\theta(k_{0}-\epsilon)e^{-ikx}a_{\mu}(k), \nonumber\\
B_{\mu \nu}^{(+)}(x)=\frac{1}{2 \pi}\int d^{3}k~
\theta(k_{0})e^{-ikx}b_{\mu \nu}(k), \label{annihilation op.}
\end{eqnarray}
and the creation operators are defined by the hermitian conjugate of
(\ref{annihilation op.}). 
$\epsilon$ in the definitions (\ref{annihilation op.}) 
is an infra-red cut-off parameter for the dipole
fields\cite{Nakanishi2, Nakanishi3}.
 
The commutation relations of the operators can be calculated by
the three-dimensional commutation relations 
(\ref{three-dimensional commutators}).
The non-vanishing commutation relations are
\begin{eqnarray}
\left[b(p),a_{\mu}^{+}(k)\right]\hspace{-0.2cm}&=&\hspace{-0.2cm}
\left[c(p),c_{\mu}^{+}(k)\right]=i p_{\mu}
\theta(p_{0}) \delta(p^{2}) \delta^{3}(p-k), \nonumber\\
\left[a_{\mu}(p),a_{\nu}^{+}(k)\right]\hspace{-0.2cm}&=&\hspace{-0.2cm}
\left[c_{\mu}(p),a_{\nu}^{+}(k)\right]=-\eta_{\mu \nu} \theta(p_{0}) 
\delta(p^2) \delta^{3}(p-k) -
p_{\mu}p_{\nu} \theta(p_{0}) \delta^{'}(p^2) \delta^{3}(p-k), \nonumber\\
\left[c_{\mu}(p),b_{\nu \rho}^{+}(k)\right]\hspace{-0.2cm}&=&\hspace{-0.2cm}
i (\eta_{\mu \nu}p_{\rho}-\eta_{\mu \rho}p_{\nu})
\theta(p_{0}) \delta(p^2) \delta^{3}(p-k), \nonumber\\
\left[b_{\mu \nu}(p),b_{\rho\sigma}^{+}(k)\right]\hspace{-0.2cm}&=&
\hspace{-0.2cm}\bigl(-\eta_{\mu\rho}p_{\nu}p_{\sigma}+\eta_{\mu\sigma}p_{\nu}
p_{\rho}+\eta_{\nu\rho}p_{\mu}p_{\sigma}-\eta_{\nu\sigma}p_{\mu}p_{\rho}
\nonumber\\
\hspace{-0.2cm}&&\hspace{-0.3cm}+~m^{2}\left(\eta_{\mu\rho}\eta_{\nu\sigma}
-\eta_{\mu\sigma}\eta_{\nu\rho}
\right)\bigr)\theta(p_{0})\delta(p^2-m^2)
\delta^{3}(p-k)\nonumber\\
\hspace{-0.2cm}&&\hspace{-0.3cm}
-\bigl(-\eta_{\mu\rho}p_{\nu}p_{\sigma}+\eta_{\mu\sigma}p_{\nu}
p_{\rho}+\eta_{\nu\rho}p_{\mu}p_{\sigma}-\eta_{\nu\sigma}p_{\mu}p_{\rho}
\bigr)\theta(p_{0})\delta(p^{2})\delta^{3}(p-k).
\label{commutator of op.}
\end{eqnarray}
In terms of these operators, the physical state conditions
(\ref{phys. cond. of field}) become
\begin{eqnarray}
c_{\mu}(p)|\textrm{phys}\rangle=0, \quad b(p)|\textrm{phys}\rangle=0, 
\quad c(p)|\textrm{phys}\rangle=0. \label{phys. cond. of op.}
\end{eqnarray}

The vacuum state is defined by
\begin{eqnarray}
b_{\mu \nu}(p)|\textrm{vac}\rangle\hspace{-0.2cm}&=&\hspace{-0.2cm}0,\quad 
a_{\mu}(p)|\textrm{vac}\rangle=0, \\
c_{\mu}(p)|\textrm{vac}\rangle\hspace{-0.2cm}&=&\hspace{-0.2cm}0,\quad 
b(p)|\textrm{vac}\rangle=0,\quad c(p)|\textrm{vac}\rangle=0. 
\end{eqnarray}
The vacuum state is physical by definition.
One particle states are constructed by the creation operators 
from the vacuum state. Physical one particle states, which
satisfy the conditions (\ref{phys. cond. of op.}), are constructed
by the creation operators which commute with $c_{\mu}(p)$, $b(p)$, and
$c(p)$. The physical states are summarized as follows\footnote{
$f_{\mu \nu}(p)$ and $c_{\mu \nu}(p)$ are the Fourier transforms of 
$F_{\mu \nu}^{(+)}(x:\epsilon)$ and $C_{\mu \nu}^{(+)}(x:\epsilon)$, 
respectively.}.
\def\labelenumi{(\roman{enumi})}
\begin{enumerate}
\item Physical massless states in the momentum frame $p_{\mu} = (p,0,p)$, 
      \begin{eqnarray}
      c_{1}^{+}(p)|\textrm{vac}\rangle, \quad
      \left(c_{0}^{+}(p)-c_{2}^{+}(p)\right)|\textrm{vac}\rangle, \quad
      b^{+}(p)|\textrm{vac}\rangle. \label{massless state}
      \end{eqnarray}
\item Physical massive states with mass $m$ 
      in the rest frame $p_{\mu} = (m,0,0)$,
      \begin{eqnarray}
      u_{\mu \nu}^{+}(p)|\textrm{vac}\rangle \equiv 
      \left(b_{\mu \nu}^{+}(p)-f_{\mu \nu}^{+}(p)+c_{\mu \nu}^{+}(p)\right)
      |\textrm{vac}\rangle.
      \end{eqnarray}      
\end{enumerate}
Here $u_{\mu \nu}^{+}(p)$ is the creation operator of the gauge invariant
field $U_{\mu\nu}(x)\equiv B_{\mu\nu}(x)-F_{\mu\nu}(x)+C_{\mu\nu}(x)$.
Note that the massless states of photon $a_{\mu}(p)$ are all unphysical
due to the ``large'' gauge symmetry with 1-form gauge function 
(\ref{gauge sym.}). 

One can show that all the physical massless states (\ref{massless
state}) are null states from the commutation relations
(\ref{commutator of op.}). The physical massive state 
$u_{12}^{+}(p)|\textrm{vac}\rangle$ is the propagating states with
positive norm and $u_{01}^{+}(p)|\textrm{vac}\rangle$ and 
$u_{02}^{+}(p)|\textrm{vac}\rangle$ are null states in the rest frame.
Thus we conclude that the physical propagating degree of
freedom\footnote{In this paper, the physical propagating state 
means the physical state with positive norm 
which contributes to the physical amplitudes.}
of the model is a physical massive state $u_{12}^{+}(p)|\textrm{vac}\rangle$
with mass $m$. Although the action (\ref{gauge inv. action}) has
the gauge symmetry (\ref{gauge sym.}), the physical massive state
appears through the generalized Stueckelberg
formalism, which is
the anti-symmetric tensor field version of the Stueckelberg formalism of
QED\cite{Stueckelberg}. 
The anti-symmetric tensor field $B_{\mu \nu}$ ``eats'' the
degrees of freedom of the gauge field $A_{\mu}$ and become a massive
anti-symmetric tensor field.
Note that massless second rank anti-symmetric
tensor field has no physical propagating degrees of freedom and 
massive one has one physical propagating degree in (2+1)-dimension
\footnote{Massless photon has also one physical propagating
degree in (2+1)-dimension.}.   
 
We consider the interesting limit of the model, $m\rightarrow 0$.
This corresponds to the limit where the
modes of closed string decouple in the corresponding effective action 
of string theory discussed in the previous subsection.
In this limit, the commutation relation of $B_{\mu \nu}$ in 
(\ref{three-dimensional commutators}) becomes
\begin{eqnarray}
\left[B_{\mu \nu}(x),B_{\rho \sigma}(y)\right] = 0,
\end{eqnarray} 
and the commutators of the creation and annihilation operators also
become
\begin{eqnarray}
\left[b_{\mu \nu}(p),b_{\rho \sigma}^{+}(k)\right] = 0.
\end{eqnarray}
Thus the states associated with 
the anti-symmetric tensor field $B_{\mu \nu}$ become zero norm.
In this limit, the physical propagating massless state in the momentum
frame $p_{\mu}=(p,0,p)$ is given by
\begin{eqnarray}
\left(u_{01}^{+}(p)-u_{12}^{+}(p)\right)|\textrm{vac}\rangle.
\label{physical propagating state}
\end{eqnarray}
Indeed, the norm of this physical propagating state becomes
\begin{eqnarray}
\hspace{-0.7cm}\Bigl\langle 
\bigl(u_{01}(p)-u_{12}(p)\bigr)\bigl(u_{01}^{+}(p)-
u_{12}^{+}(p)\bigr)\Bigr\rangle &=& \Bigl\langle 
\bigl(f_{01}(p)-f_{12}(p)\bigr)\bigl(f_{01}^{+}(p)-f_{12}^{+}(p)\bigr)
\Bigr\rangle \nonumber \\ 
&=& 4 p^{2} \Bigl\langle
a_{1}(p)~a_{1}^{+}(p)\Bigr\rangle.
\end{eqnarray}  
Hence, in this limit, the physical propagating state becomes essentially the
transverse photon $a_{1}^{+}(p)|\textrm{vac}\rangle$.
However it is worth noting that even though the norms of 
$\left(u_{01}^{+}(p)-u_{12}^{+}(p)\right)|\textrm{vac}\rangle$ and 
$\left(f_{01}^{+}(p)-f_{12}^{+}(p)\right)|\textrm{vac}\rangle$ are same in the
limit $m\rightarrow 0$, 
the physical propagating state is \textit{not} 
$\left(f_{01}^{+}(p)-f_{12}^{+}(p)\right)|\textrm{vac}\rangle$, 
\textit{but} $\left(u_{01}^{+}(p)-u_{12}^{+}(p)\right)|\textrm{vac}\rangle$.

This situation is similar to the broken phase of Yang-Mills-Higgs
model, where the equivalence theorem holds\cite{Gaillard, Aoki}. 
This theorem claims the amplitude for emission or absorption of 
the longitudinal states of 
the massive gauge boson becomes equal, at high energy, to 
the amplitude for emission or absorption of
the unphysical Nambu-Goldstone states, which is
``eaten'' by the gauge boson.
In our model, the physical massive state of the anti-symmetric tensor field
appears after the anti-symmetric tensor field 
``eats'' the unphysical state of the transverse photon.  
The above analysis of the norm of the physical states implies 
that the analogous equivalence theorem holds in our model:
in the high energy region where one can ignore mass $m$, the amplitude
for emission or absorption of the longitudinal states of the 
physical massive anti-symmetric tensor field
is the same as the amplitude for emission or absorption of 
the unphysical massless transverse photon.
   
\section{Spontaneous Lorentz symmetry breaking by anti-symmetric
tensor field}
In this section we discuss the spontaneous breaking of the
Lorentz symmetry by a constant vacuum expectation value (vev) of
the second rank anti-symmetric tensor field in our model.

\subsection{The Nambu-Goldstone boson for the spontaneous Lorentz 
symmetry breaking}
The equations of motion (\ref{eq. of B})-(\ref{eq. of NLC}) have
a solution such that only $B_{\mu \nu}$ and $F_{\mu \nu}$ have constant
nonzero vev's\footnote{If the vev $\langle\tilde{B}^{\mu}\rangle~\left(\equiv
\frac{1}{2}\epsilon^{\mu \nu \rho} \langle B_{\nu \rho}\rangle\right) 
=\langle\tilde{F}^{\mu}\rangle$ is a time-like constant vector, 
one can transform it to this form by an appropriate Lorentz
transformation.}:
\begin{eqnarray}
\langle B_{12} \rangle = \langle F_{12} \rangle = B_{\textrm{vev}} 
= \textrm{const.}~(\neq 0), \label{breaking vacuum} 
\end{eqnarray}      
One can easily find that this solution is a ground state with
vanishing energy of the Hamiltonian derived from the action
(\ref{total action}). (See (\ref{energy-momentum tensor}).)
Although $B_{\textrm{vev}}$ is an undetermined constant
in our model, we assume this to be a nonzero constant in the
sequel. Since the vev's of $C_{\mu}$, $B$, and $C$ vanish, 
this vacuum is a physical state. 
In frameworks of the nonperturbative string theory, the possibility of 
$\langle B_{\mu \nu} \rangle \neq 0$ has been discussed in
\cite{Samuel1, Samuel2}.
In this viewpoint, the nonzero vev of $B_{12}$ induces a spontaneous
magnetization $\langle F_{12} \rangle$.

Since $B_{\mu \nu}$ and $F_{\mu \nu}$ are not gauge invariant under the gauge
transformation (\ref{gauge sym.}), one may expect to be able to eliminate
this vev by the gauge transformation. However the gauge transformation
which eliminates the vev requires the linear 1-form gauge functions such
as $\Lambda_{1}(x)=\frac{1}{2}B_{\textrm{vev}}x^{2}$ and 
$\Lambda_{2}(x)=-\frac{1}{2}B_{\textrm{vev}}x^{1}$.
These gauge functions are ill-defined at the infinity. Hence we do not
require the invariance under such singular gauge transformations 
to define the Hilbert space of the quantum theory. 
%as in the case of the non-abelian gauge theory [].  

As discussed in \cite{Yokoi}, on the vacuum (\ref{breaking vacuum}), 
the (2+1)-dimensional Lorentz symmetry $SO(2,1)\sim SL(2,\bf{R})$ 
is spontaneously broken down to the spatial rotation $SO(2)\sim U(1)$ 
by the vev of $B_{12}$ and $F_{12}$
\footnote{Our convention for the Poincar\'e algebra 
in (2+1)-dimension is
\begin{eqnarray*}
&&\left[P_{\mu}, P_{\nu}\right]=0, \quad \left[M_{\mu \nu}, P_{\rho}\right] =
-i \left(\eta_{\mu \rho}P_{\nu}-\eta_{\nu
\rho}P_{\mu}\right), \\
&&\left[M_{\mu \nu}, M_{\rho \sigma}\right] = -i\left(\eta_{\mu
\rho}M_{\nu \sigma}-\eta_{\nu \rho}M_{\mu \sigma}+\eta_{\nu \sigma}
M_{\mu \rho}-\eta_{\mu \sigma}M_{\nu \rho}\right).  
\end{eqnarray*} \label{Poincare}}. 
What is the Nambu-Goldstone (NG) bosons for the broken boost
generators in this model ?

In order to answer this question, we construct the generators 
of the Lorentz transformations $M_{\rho \sigma}$ by the Noether method.
The conserved currents $M_{\rho \sigma}{}^{\mu}(x)$ for 
the Lorentz symmetry can be derived from the action (\ref{total
action})\footnote{$\left(S_{\rho \sigma}\phi\right)_{I}$ 
denotes infinitesimal transformations of internal spin: 
$\left(S_{\rho \sigma}\chi\right)=0$ for scalar fields,
$\left(S_{\rho \sigma}V\right)_{\mu}=i \left(\eta_{\rho
\mu}V_{\sigma}-\eta_{\sigma \mu}V_{\rho}\right)$ for vector fields,
and $\left(S_{\rho \sigma}B\right)_{\mu \nu} = i \left(\eta_{\rho
\mu}B_{\sigma \nu}-\eta_{\sigma \mu}B_{\rho \nu}+\eta_{\rho
\nu}B_{\mu \sigma}-\eta_{\sigma \nu}B_{\mu \rho}\right)$ for second
rank anti-symmetric tensor fields.}:
\begin{eqnarray}
M_{\rho \sigma}{}^{\mu}(x)\hspace{-0.2cm}&=&\hspace{-0.2cm} 
x_{\rho}T_{\sigma}{}^{\mu}(x)-x_{\sigma}T_{\rho}{}^{\mu}(x) -i
\frac{\partial {\cal L}_{\textrm{total}}(x)}
{\partial (\partial_{\mu} \phi_{I})}\left(S_{\rho
\sigma}\phi\right)_{I}\nonumber\\
\hspace{-0.2cm}&=&\hspace{-0.2cm}\frac{1}{2 m^2}H^{\mu \alpha \beta}
\left(x_{\rho}\partial_{\sigma}-x_{\sigma}\partial_{\rho}\right)
B_{\alpha \beta} - \left(F^{\mu \alpha}-B^{\mu \alpha}\right) \left(x_{\rho}
\partial_{\sigma}-x_{\sigma}\partial_{\rho}\right) A_{\alpha}
\nonumber\\
\hspace{-0.2cm}&&\hspace{-0.2cm}+\,C_{\alpha}
\left(x_{\rho}\partial_{\sigma}-x_{\sigma}
\partial_{\rho}\right)B^{\mu \alpha}
-B\left(x_{\rho}\partial_{\sigma}-x_{\sigma}\partial_{\rho}\right)A^{\mu}
-C\left(x_{\rho}\partial_{\sigma}-
x_{\sigma}\partial_{\rho}\right)C^{\mu} \nonumber\\
\hspace{-0.2cm}&&\hspace{-0.2cm}-\,\left(x_{\rho}\delta^{\mu}_{\sigma}-
x_{\sigma}\delta^{\mu}_{\rho}\right)\Bigl(\frac{1}{12 m^2}
\left(H_{\alpha \beta \gamma}\right)^{2}-\frac{1}{4}\left(F_{\alpha
\beta}-B_{\alpha \beta}\right)^{2}+C^{\beta}\partial^{\alpha}B_{\alpha
\beta}\Bigr) \nonumber\\
\hspace{-0.2cm}&&\hspace{-0.2cm}+\,\Bigl(\frac{1}{m^2}
H^{\mu \alpha \beta}+\eta^{\mu
\alpha}C^{\beta}-\eta^{\mu \beta}C^{\alpha}\Bigr)\left(\eta_{\rho
\alpha}B_{\sigma \beta}-\eta_{\sigma \alpha}B_{\rho \beta}\right)
\nonumber\\
\hspace{-0.2cm}&&\hspace{-0.2cm}-\,\left(F^{\mu \alpha}-
B^{\mu \alpha}+\eta^{\mu \alpha}B\right)
\left(\eta_{\rho \alpha}A_{\sigma}-\eta_{\sigma \alpha}A_{\rho}\right)
-\eta^{\mu \alpha}C \left(\eta_{\rho \alpha}C_{\sigma}
-\eta_{\sigma \alpha}C_{\rho}\right),
\label{Lorentz conserved current}
\end{eqnarray}
where the canonical energy-momentum tensor $T_{\rho}{}^{\mu}(x)$ is given
by
\begin{eqnarray}
T_{\rho}{}^{\mu}(x)\hspace{-0.2cm}&=&\hspace{-0.2cm}\frac{1}{2 m^2}
H^{\mu \alpha
\beta}\partial_{\rho}B_{\alpha \beta}-\left(F^{\mu \alpha}-B^{\mu
\alpha}\right)\partial_{\rho}A_{\alpha}\nonumber \\
\hspace{-0.2cm}&&\hspace{-0.2cm}+\,C_{\alpha}\partial_{\rho}
B^{\mu \alpha}-B\partial_{\rho}A^{\mu}
-C\partial_{\rho}C^{\mu} \nonumber\\
\hspace{-0.2cm}&&\hspace{-0.2cm}-\,\delta^{\mu}_{\rho}\Bigl(\frac{1}{12 m^2}
\left(H_{\alpha \beta \gamma}\right)^{2}-\frac{1}{4}\left(F_{\alpha
\beta}-B_{\alpha \beta}\right)^{2}+C^{\beta}\partial^{\alpha}B_{\alpha
\beta}\Bigr).\label{energy-momentum tensor}
\end{eqnarray}
From the conserved currents (\ref{Lorentz conserved current}), one can
obtain the generators of Lorentz transformation $M_{\rho \sigma}$:
\begin{eqnarray}
M_{\rho \sigma}=\int d^{2}{\boldsymbol x} M_{\rho \sigma}{}^{0}(x).
\label{Noether charge}
\end{eqnarray}

By utilizing the expressions (\ref{Lorentz conserved current}) and 
(\ref{Noether charge}) and the commutation relations
(\ref{three-dimensional commutators}), we have the 
nonvanishing vev of the following
commutation relations on the vacuum (\ref{breaking vacuum}):
\begin{eqnarray}
\bigl\langle\left[i M_{0i}, B_{0j}(x)\right]\bigr\rangle
&=&\int\!d^{2}{\boldsymbol y} \bigl\langle[i M_{0i}{}^{0}(y), B_{0j}(x)]
\bigr\rangle = \epsilon_{ij}B_{\textrm{vev}},\\
\bigl\langle\left[i M_{0i}, F_{0j}(x)\right]\bigr\rangle 
&=&\int\!d^{2}{\boldsymbol y} \bigl\langle[i M_{0i}{}^{0}(y), F_{0j}(x)]
\bigr\rangle = \epsilon_{ij}B_{\textrm{vev}}\quad(\epsilon_{12}=1). 
\end{eqnarray}
Thus two boost generators $M_{0i}~(i=1,2)$ are spontaneously broken on
the vacuum. From the above commutation relations, the 
candidates for the NG-bosons for the broken boost generators are
$B_{0i}$ and $F_{0i}$. However, $B_{\mu \nu}$ is a mixture of massive 
and massless components as discussed previously.
As obtained in the previous section, the mass eigenstates of the model
are the massive field $U_{\mu \nu}$ and the massless field $F_{\mu \nu}$ (or
$A_{\mu}$) and the other massless fields $C_{\mu}$, $B$, and $C$.
Since NG-bosons are necessarily massless state, we conclude that 
the NG-bosons for the broken boost generators are the massless photon
$F_{0i}$. Incidentally, the massive field $U_{\mu \nu}$ satisfies
\begin{eqnarray}
\bigl\langle\left[i M_{0i}, U_{\mu \nu}(x)\right]\bigr\rangle = 
\int\!d^{2}{\boldsymbol y} \bigl\langle[i M_{0i}{}^{0}(y), U_{\mu \nu}(x)]
\bigr\rangle = 0.
\end{eqnarray}
Thus the fact that $B_{\mu \nu}$ becomes essentially a massive field 
is consistent with the Nambu-Goldstone theorem.

\subsection{Are the NG-bosons physical ?}
We have identified the NG-bosons for the spontaneous Lorentz symmetry
breaking with the massless photon $F_{0i}$.
The one particle state of the massless 
photon $F_{\mu \nu}$ (or $A_{\mu}$) 
is an unphysical state as discussed in the
previous section.
Hence the NG-bosons for the spontaneous Lorentz symmetry breaking 
are unphysical states, i.e.,
unphysical NG-bosons.
Since the Lorentz symmetry is a physical global symmetry, 
the corresponding NG-bosons are expected to be physical. 
Is this a contradiction ?

In order to answer this question, let us discuss 
the one particle state created by the broken
boost generators from the vacuum (\ref{breaking vacuum}).
Inserting the decompositions $\phi_{I}(x) = \langle 
\phi_{I} \rangle + \hat{\phi}_{I}(x)$ on the vacuum\footnote{
Since $\langle F_{12}\rangle = B_{\textrm{vev}}$, we take, for an example, 
the vev's of $A_{i}$ as $\langle A_{1} \rangle = -\frac{1}{2} 
B_{\textrm{vev}} x^{2}$ and 
$\langle A_{2} \rangle = \frac{1}{2} B_{\textrm{vev}} x^{1}$.} 
into the expressions of 
the currents (\ref{Lorentz conserved current}), 
one can construct the broken parts of the currents 
$M_{0i}^{B}{}^{\mu}(x)$ which depends on $B_{\textrm{vev}}$, i.e., 
$M_{0i}{}^{\mu}=M_{0i}^{B}{}^{\mu}(B_{\textrm{vev}},\hat{\phi}_{I})+
\hat{M}_{0i}{}^{\mu}(\hat{\phi}_{I})$.
The explicit forms of the broken parts of the 
currents $M_{0i}^{B}{}^{\mu}(x)$ are 
\begin{eqnarray}
M_{0i}^{B}{}^{\mu}(x) &=& B_{\textrm{vev}}~\epsilon_{ij} \biggl\{
\frac{1}{2}x^{j}\left(\hat{F}^{\mu 0}-\hat{B}^{\mu 0}+\eta^{\mu 0}
\hat{B} \right)
- \frac{1}{2}x_{0}\left(\hat{F}^{\mu j}-\hat{B}^{\mu j}+\eta^{\mu j} 
\hat{B}\right)
\biggr.
\nonumber \\ 
&& \biggr. +\Bigl(\frac{1}{m^2}\hat{H}^{\mu 0 j} + 
\eta^{\mu 0}\hat{C}^{j}-\eta^{\mu j} \hat{C}^{0}\Bigr)\biggr\}
\label{broken current}
\end{eqnarray}
and $M_{12}^{B}{}^{\mu}$ vanishes. 
%Without confusion, we abbreviate simply $\hat{\phi}_{I}$ as $\phi_{I}$.
Using the equations of motion for $\hat{\phi}_{I}$ which are the same as  
(\ref{eq. of B})-(\ref{eq. of NLC}), one can
easily show that these currents are also conserved.
From these currents one can obtain the broken parts of 
the generators which are conserved:
\begin{eqnarray}
M_{0i}^{B}&\equiv&\int d^{2}{\boldsymbol x}~M_{0i}^{B}{}^{0}(x)
\nonumber \\&=& \hspace{-0.1cm}B_{\textrm{vev}}\!
\int \! d^2{\boldsymbol x}~\epsilon_{ij}\Bigl\{
\bigl(\frac{1}{2}x_{0}\hat{U}^{0j}(x)\bigr)+\bigl(-\frac{1}{2}x_{0}
\hat{C}^{0j}(x) +
\frac{1}{2}x^{j}\hat{B}(x)+\hat{C}^{j}(x)\bigr)\Bigr\}.
\label{broken generator}
\end{eqnarray}
Hereafter we abbreviate simply $\hat{\phi}_{I}$ as $\phi_{I}$ without
confusion. These broken parts of generators indeed satisfy
\begin{eqnarray}
\left[iM_{0i}^{B},B_{0j}(x)\right]=
\left[iM_{0i}^{B},F_{0j}(x)\right]
= \epsilon_{ij} B_{\textrm{vev}}. \label{commutators of broken generator}
\end{eqnarray}
Note that $M_{0i}^{B}$ is a physical operator which commutes with
$C_{\mu}$, $B$, and $C$.

We consider the one particle state created by $M_{0i}^{B}$ from the vacuum
(\ref{breaking vacuum}), denoted as $|\textrm{VAC}\rangle$.
Since we are interested in the NG-bosons for the broken generators,
we consider only the massless state of the one particle state given by  
\begin{eqnarray}
M_{0i}^{B}\bigl|\textrm{VAC}\bigr\rangle\Bigr|_{\textrm{massless}}
= B_{\textrm{vev}} \! \int \! d^{2}{\boldsymbol x}~\epsilon_{ij}
\bigl(-\frac{1}{2}x_{0} C ^{0j}(x) +
\frac{1}{2}x^{j} B(x) + C^{j}(x)\bigr)\bigl|\textrm{VAC}\bigr\rangle.
\label{massless state of broken generator}
\end{eqnarray}
%The state which gives the commutators (\ref{commutator of broken
%generator}) is essentially the third term in 
%(\ref{massless state of broken generator}). 
From the commutation relations (\ref{three-dimensional commutators}),
one can easily find that these states are the physical states which
satisfy the conditions (\ref{phys. cond. of field}), but the null states. 
Therefore we conclude that although the spontaneous Lorentz symmetry 
breaking is \textit{physical}, 
the massless one particle state created by the broken generator
$M_{0i}^{B}$ from the vacuum (\ref{breaking vacuum}) 
is not only a physical state but also \textit{null} state. 
Since our model has the twisted structure of the Hilbert space 
with an indefinite metric 
on account of the gauge invariance (\ref{gauge sym.}), 
nonzero matrix elements exist between physical null states 
and unphysical states.
Thereby the NG-boson which has the nonzero matrix element with
this massless one particle state can be \textit{unphysical}. 
This is closely analogous to the abelian Higgs model 
in the NL formalism \cite{Nakanishi2}.

As discussed in the previous section, in the limit $m\rightarrow 0$,
the unphysical transverse photon, which is the 
unphysical NG-boson for the spontaneous Lorentz symmetry
breaking, becomes essentially the physical propagating state of the model. 
The implication of this fact will be discussed in the next section.  

\section{A perturbation}
So far, we have studied the free field theory of $B_{\mu \nu}$ and
$A_{\mu}$ whose action is given by (\ref{total action}).
In this section, we discuss a possible perturbation of the model. 
We introduce the following interaction terms to (\ref{total action})
as the perturbation:
\begin{eqnarray}
{\cal L}_{\textrm{int}}(x)= \sum_{n=2}^{N} a_{n} 
\bigl\{\left(F_{\mu\nu}-B_{\mu\nu}\right)^2\bigr\}^{n}.
\label{interaction}
\end{eqnarray}
This type of interactions is obtained by the $\alpha^{'}$-expansion 
of DBI action\footnote{By virtue of the peculiarity of (2+1)-dimension,
a Lorentz scalar constructed from $F_{\mu \nu}-B_{\mu \nu}$ can be 
always expressed as a polynomial of $\left(F_{\mu \nu}-
B_{\mu \nu}\right)^{2}$.}.
These interactions are gauge invariant under 
the gauge transformation (\ref{gauge sym.}) and consistent 
with the physical state conditions (\ref{phys. cond. of field}):
these interaction terms do not change the equations of motion of 
$C_{\mu}$, $B$, and $C$ in (\ref{free eq. of motion}) 
and keep them free fields.
Although these interactions are non-renormalizable,
we treat them as the perturbations which are interpreted as the 
operator insertions in the matrix elements 
in the similar manner to the chiral Lagrangian of QCD.

Although the equations of motion of $B_{\mu \nu}$ and $A_{\mu}$ is 
modified due to the interaction terms, the solution (\ref{breaking
vacuum}) still remains to be a solution of the modified equations.
Thus, even after including the perturbations, the spontaneous Lorentz
symmetry breaking is realized. In this case, the argument about 
the NG-boson in the previous section does not change essentially 
and hence we can conclude that the NG-boson remains to be the
unphysical photon. 

As is well-known, the low-energy dynamics of the 
NG-boson is given by only the symmetry argument, i.e., the low-energy
theorem. In particular, the low-energy effective action of the
NG-boson is given by the nonlinear realization of the broken
symmetry. In the case of the spontaneous Lorentz symmetry breaking,
these argument holds and the low-energy effective action of the
NG-boson is expected to be given by the nonlinear realization of the
broken Lorentz symmetry.
This problem will be argued in the next section. 

In the limit $m\rightarrow 0$, the physical 
propagating state becomes essentially the unphysical transverse photon.
Furthermore the amplitude of the physical propagating state
(\ref{physical propagating state}), for example, the 
two-point amplitude satisfies  
\begin{eqnarray}
&& \hspace{-3cm}\Bigl\langle \left(u_{01}(p)-u_{12}(p)\right)
\tilde{{\cal L}}_{\textrm{int}}(q)\left(u_{01}^{+}(k)-
u_{12}^{+}(k)\right)\Bigr\rangle \nonumber\\
&& = \Bigl\langle \left(f_{01}(p)
-f_{12}(p)\right)\tilde{{\cal L}}_{\textrm{int}}^{(b_{\mu \nu}=0)}(q)
\left(f_{01}^{+}(k)-f_{12}^{+}(k)\right)
\Bigr\rangle,
\end{eqnarray}
where $\tilde{{\cal L}}_{\textrm{int}}(q)$ and 
$\tilde{{\cal L}}_{\textrm{int}}^{(b_{\mu \nu}=0)}(q)$ are
the Fourier transforms of ${\cal L}_{\textrm{int}}(x)$ and 
${\cal L}_{\textrm{int}}^{(B_{\mu \nu}=0)}(x)$ which is obtained by
setting $B_{\mu \nu}=0$ in (\ref{interaction}), respectively.
This relation can be generalized to the scattering amplitudes for any
number of the incoming or outgoing physical particles.
Thus the physical amplitude in this limit is given by 
the amplitude of the unphysical transverse photon, that is, 
the NG-boson of the spontaneous Lorentz symmetry breaking.
This is an analogue of the equivalence theorem of 
the Yang-Mills-Higgs model. In the energy region 
$(B_{\textrm{vev}})^{\frac{2}{3}} \gg E \gg m$, where
$(B_{\textrm{vev}})^{\frac{2}{3}}$ is the scale of the Lorentz
symmetry breaking, the physical S-matrix elements can be obtained 
by the scattering amplitudes of the NG-bosons for the Lorentz
symmetry breaking. 

\section{Discussions}
In this paper, we have studied a covariant canonical quantization 
of a gauge invariant model of a second rank 
anti-symmetric tensor field and a vector field (photon). 
The spontaneous Lorentz symmetry breaking on the vacuum with 
a constant vev of the anti-symmetric tensor field has also been
studied and the NG-boson of the Lorentz symmetry breaking has been
identified with the unphysical photon.
In this section, we discuss some related problems.
\subsection{The spontaneous symmetry breaking of translation}
Until now, we have discussed only the spontaneous Lorentz symmetry
breaking on the vacuum (\ref{breaking vacuum}).
Indeed, the vacuum breaks the translational symmetry, 
because the vev $\langle F_{12}\rangle = B_{\textrm{vev}}$
leads to, for example, the vev's
\begin{eqnarray}
\langle A_{1}\rangle = -\frac{1}{2}B_{\textrm{vev}} x^{2} \quad
\textrm{and}\quad
\langle A_{2}\rangle = \frac{1}{2}B_{\textrm{vev}} x^{1}.
\label{translation breaking vacuum}
\end{eqnarray}
Since $A_{1}$ and $A_{2}$ are not gauge invariant,  
one may expect that the vev's can be eliminated by a gauge transformation.
However, following the discussion about the vev's 
of $B_{\mu \nu}$ and $F_{\mu \nu}$ in section 3, 
we do not require the invariance of the Hilbert space 
under the singular gauge transformation which eliminates them. 

By the vev's (\ref{translation breaking vacuum}), two translation
generators $P_{1}$ and $P_{2}$ are broken:
\begin{eqnarray}
\bigl\langle \left[ i P_{i}, A_{j}(x)\right] \bigr\rangle = \epsilon_{ij}
\frac{1}{2}B_{\textrm{vev}},
\end{eqnarray}
where $P_{i}$ is given by the canonical energy-momentum tensor
(\ref{energy-momentum tensor})
\begin{eqnarray}
P_{i}=\int d^{2} {\boldsymbol x}~T_{i}^{0}(x).
\end{eqnarray}
The NG-bosons associated with the broken translation generator $P_{1}$
and $P_{2}$ are $A_{2}$ and $A_{1}$ respectively.
The similar discussion to the broken Lorentz symmetry concludes that 
the NG-boson of the broken translational symmetry is also
the unphysical massless photon.

This can be also understood from the following 
commutation relation in the Poincar\'e algebra\footnoteref{Poincare}:
\begin{eqnarray}
\left[M_{0i}, P_{0}\right] = -i P_{i}.
\end{eqnarray}
Sandwiching the Jacobi identity 
\begin{eqnarray}
\bigl[\left[M_{0i}, P_{0}\right], A_{j}(x)\bigr] +
\bigl[\left[A_{j}(x), M_{0i}\right], P_{0}\bigr] +
\bigl[\left[P_{0}, A_{j}(x)\right], M_{0i}\bigr] = 0
\end{eqnarray} 
between the vacuum states $|\textrm{VAC}\rangle$, 
one can obtain the following equality
\begin{eqnarray}
\bigl\langle\left[iP_{i}, A_{j}(x)\right]\bigr\rangle = 
\bigl\langle\left[i M_{0i}, \partial_{0} A_{j}(x)\right]\bigr\rangle
=\epsilon_{ij}\frac{1}{2}B_{\textrm{vev}},
\end{eqnarray}
where we have used $P_{0}|\textrm{VAC}\rangle = 0$ and 
$\left[P_{0}, \phi_{I}\right] = -i \partial_{0} \phi_{I}$.
This implies that when the NG-boson of the broken translation
generator $P_{i}$ is $A_{j}$, the NG-boson of the broken Lorentz
generator $M_{0i}$ is given by its time derivative 
$\partial_{0}A_{j} \sim F_{0j}$.     
This phenomenon has been known as the inverse Higgs phenomenon
in the nonlinear realization of space-time symmetries\cite{Ivanov1}.

\subsection{Relation to the nonlinear realization of Lorentz symmetry}
According to the discussion in the previous subsection,
the low-energy effective action of the NG-boson $A_{i}$ can be
obtained by the nonlinear realization of the translational and
Lorentz symmetry, which leads to a 1-dimensional effective action
\footnote{The Nambu-Goto type effective action on a 
lower-dimensional brane embedded in higher dimensional 
flat space-time is known to be obtained by the nonlinear realization 
of the higher-dimensional translational and Lorentz symmetries
\cite{Ivanov2}.}.  
However, since no physical lower-dimensional object 
such as a brane exist in our model, we expect a (2+1)-dimensional
effective action which describes the low-energy effective theory in
the whole space-time.

To realize this expectation, as in the case of the broken boost
generators, we split the broken 
translation generators $P_{i}$ into the broken parts and the
unbroken parts such as
$P_{i}=P_{i}^{B}(B_{\textrm{vev}})+\hat{P}_{i}$
\footnote{As in the case of the boost generators, one can show that 
$P_{i}^{B}$ and $\hat{P}_{i}$ are conserved separately.}:
\begin{eqnarray}
P_{i}^{B}=\frac{B_{\textrm{vev}}}{2} \int d^{2} {\boldsymbol x}~
\epsilon_{ij}\left(\hat{U}^{0j}-\hat{C}^{0j}\right).
\end{eqnarray}
The broken parts of the translation generators satisfy
\begin{eqnarray}
\left[i P_{i}^{B}, A_{j}\right] = \epsilon_{ij}
\frac{1}{2}B_{\textrm{vev}}.
\end{eqnarray}
The same argument as the case of the boost generators leads us to the
conclusion that the NG-bosons for the broken translation generators 
can be the unphysical massless photon.

Here, if one considers \textit{only} the physical Hilbert space
defined by the physical state conditions (\ref{phys. cond. of field}),
one can show that the generators $\{ M_{\mu \nu},~P_{0},~\hat{P}_{i}\}$
%\footnote{$P_{i}^{B}$ essentially decouples (?).} 
form a closed Poincar\'e algebra on the physical Hilbert space.
As far as one considers this Poincar\'e algebra on the physical
Hilbert space, the translational symmetry generated by $\hat{P}_{i}$ is
\textit{unbroken} and only the boost symmetry generated by $M_{0i}$ 
is broken. In this breaking pattern of the Poincar\'e symmetry, 
the low-energy effective action constructed by the nonlinear realization is
a (2+1)-dimensional effective action and its explicit form 
has been obtained in \cite{Yokoi}.
Thus, we expect that the physical amplitudes of the model 
can be obtained by the low-energy effective action 
via the equivalence theorem discussed in the previous section.  
\subsection{Other related topics} 
From the relation between the non-commutative gauge theory 
and the D-brane world volume effective theory  
on a constant background anti-symmetric tensor field, our
investigation is expected to give some new insights to the
non-commutative gauge theory.  
However, in this context, the anti-symmetric tensor field is taken as an
\textit{external} field: the kinetic term for the anti-symmetric 
tensor field does not exist. Then the vacuum 
\begin{eqnarray}
\bigl\langle B_{12} \bigr\rangle \neq 0 \quad \textrm{and} \quad
\bigl\langle F_{12} \bigr\rangle = 0,
\end{eqnarray}
is allowed as the solution of the equation of motion of
a gauge field $A_{\mu}$. 
This solution is a different vacuum 
from our vacuum (\ref{breaking vacuum}).
The extension of our analysis to the above case and the relation to
the non-commutative gauge theory are interesting problems.

We make some speculations about the 
dynamics of photon $A_{\mu}$ in our model.
At first, since the photon is the NG-boson of the broken Lorentz
symmetry on the vacuum (\ref{breaking vacuum}), the NG-theorem concludes
that the photon cannot become massive on the vacuum.
Secondly, as discussed in \cite{Yokoi}, the NG-boson of the broken Lorentz
symmetry has derivative couplings with \textit{any} fields 
including itself.
Hence, the photon is expected to have derivative coupling 
with \textit{any} fields including itself and neutral fields.
From the relation to the non-commutative gauge theory, 
this seems consistent with the fact that the non-commutative photon in 
the non-commutative QED, where the products of fields are replaced by
the $\star$-products, has self-couplings and derivative couplings 
with fields in the adjoint representation, i.e., neutral fields
\cite{Hayakawa1, Hayakawa2}.   

Although the discussion in this paper is limited to the model in 
(2+1)-dimension, the similar discussions can be applied to the models in 
higher dimensions. Investigation of the more realistic case, that is,
(3+1)-dimensional case is an interesting future problem.
  
Investigation of the supersymmetric extension of our model from the view
point in this paper is also interesting.

{\bf Acknowledgements} \\
We would like to thank N. Ohta, T. Nakatsu, M. Torii, T. Kimura 
and T. Kugo for useful discussions and comments.

\end{document}